\documentclass[aps,pra,epsfigure,twocolumn]{revtex4}
\usepackage{amsmath}
\usepackage{amstext}
\usepackage{latexsym}
\usepackage{psfig}
\usepackage{graphicx}
\usepackage{amsfonts}

\newcommand{\pro}[2]{\left\vert#1\rangle\langle#2\right\vert}
\newcommand{\ket}[1]{\left\vert#1\right\rangle}
\newcommand{\modul}[1]{\left\vert#1\right\vert}
\newcommand{\cnot} {{\textsf {CNOT}}}

\newcommand{\sprod}[2]{\left\langle#1\vert#2\right\rangle}

\newcommand{\bra}[1]{\left\langle#1\right\vert}
\newcommand{\sand}[3]{\left\langle#1\vert#2\vert#3\right\rangle}

\newcommand{\one}{\mbox{$1\hspace{-1.0mm}{\bf l}$}}
\begin{document}

\title{Vibrational coherent quantum computation}
\author{Mauro Paternostro, M. S. Kim}
\affiliation{School of Mathematics and Physics, The Queen's University, Belfast, BT7 1NN, United Kingdom}
\author{Peter L. Knight}
\affiliation{Blackett Laboratory, Imperial College London, Prince Consort Road, London SW7 2BW, United Kingdom}
\date{\today}

\begin{abstract}
A long-lived coherent state and non-linear interaction have been
experimentally demonstrated for the vibrational mode of a trapped
ion.  We propose an implementation of quantum computation using
coherent states of the vibrational modes of trapped ions. Differently
from earlier experiments, we consider a far-off resonance for the
interaction between external fields and the ion in a
bidimensional trap. By appropriate choices of the detunings
between the external fields, the adiabatic elimination of the
ionic excited level from the Hamiltonian of the system allows for
beam splitting between orthogonal vibrational modes, production of
coherent states and non-linear interactions of various kinds. In
particular, this model enables the generation of the four
coherent Bell states. Furthermore, all the necessary operations
for quantum computation such as preparation of qubits,  one-qubit
and controlled two-qubit operations, are possible. The detection
of the state of a vibrational mode in a Bell state 
is made possible by the combination of resonant and off-resonant 
interactions between the ion and some external fields. 
We show that our read-out scheme provides highly
efficient discrimination between all the four Bell states. 
We extend this to a quantum register composed of many 
individually trapped ions. In this case, operations on two remote 
qubits are possible through a cavity mode. We emphasize that our remote-qubit
operation scheme does not require a high quality factor resonator: the cavity field acts as a catalyst for the gate operation.

\end{abstract}
\maketitle
\section{Introduction}
Outstanding theoretical and experimental advances have been
reported in the field of photonic quantum information processing,
ranging from the experimental realization of the quantum
teleportation protocol~\cite{furusawa} to proposals for quantum
error correction~\cite{ralph1} and quantum computation~\cite{klm}.
It has been shown that universal continuous-variable quantum
computation can be performed using linear optics (including
squeezing), homodyne detection and non-linearities realized by
photon-counting
positive-valued-measurement~\cite{bartlettsanders}. Recently, a
method to implement efficient universal computation based on
coherent states of light has been suggested and shown to be
robust against detection inefficiencies~\cite{jacobmyung}.

As pointed out in the Los Alamos Roadmap for quantum
computing~\cite{roadmap}, using coherent states of a boson as
logical qubits is one of the promising ways to realize quantum
computation.  However, one of the practical difficulties
encountered in a scheme for coherent quantum computation is the
requirement of a strong Kerr non-linear interaction to produce a
superposition of coherent states. Currently available non-linear
dielectrics, unfortunately, offer too low rates of non-linearity
with exceedingly high absorption of the incoming field. In this
context, some recent proposals for giant Kerr non-linear
interaction exploiting electromagnetic induced transparency remains
to be proved to work in the quantum domain~\cite{ioEIT}.

In this paper, we propose to implement coherent quantum
computation using vibrational modes of trapped ions.  Since the early
days of the quantum manipulation of vibrational modes for trapped
ions, it has been clear that strong non-linear evolutions can be
efficiently engineered using two or three-level ions (in a
$\Lambda$ configuration) interacting with properly tuned laser
pulses~\cite{wineland,sasura,STK,lieb}. 
Furthermore, a long-lived coherent state has been
experimentally reported~\cite{wineland,lieb}.  This opens a way
to the exploitation of vibrational states as the elements of a
quantum register in a quantum processor. For the purposes of
scalability, arbitrarily large quantum registers have to be
considered. One way is to work with a chain of ions in the same
trap, exploiting not just the vibrational modes of the centre of
mass (CM) but the collective vibrational excitations of the chain
(see~\cite{wineland,sasura} and references within). Another way
to realize the scalability is to take advantage of the
recently demonstrated coupling between cavities and single-ion
traps~\cite{schmidt-kaler,walther}, which is the scheme used in
this proposal. In our architecture, an array of many individually
trapped ions constitute the quantum register. The local
processors are interconnected via an effective all-optical bus
realized by a cavity mode coupled to the different traps. We will
not require a perfect cavity for our protocol and the cavity
field mode never becomes entangled with the ions of the register
(the coupling between two different ions being realized via a
second-order interaction only virtually mediated by the cavity
field).

In this paper, we also propose an efficient discrimination of the
four quasi-Bell states embodied by entangled coherent
states~\cite{jacobmyung}. In this respect, our detection scheme
does not require the complete map of a quasi-Bell state onto the
discrete electronic Hilbert space of the trapped
ions~\cite{munrosanders}. The detection is performed exploiting
the additional degree of freedom of the vibrational states
represented by their even- and odd-number parities. It is worth
stressing that the Bell-state discrimination can be accomplished,
in our set-up, both locally (exploiting two orthogonal vibrational
modes of a single trapped ion) and remotely, where the Bell state
is the joint state of two vibrational modes of a linear two-ion
crystal.

The paper is organized as follows. In Section~\ref{QSE} we
describe the coupling scheme used in our proposal and address the
issue of the preparation and single-qubit manipulation of coherent
states. In this context, the generation of even/odd coherent states
and entangled coherent state is discussed. We perform some
quantitative investigations to prove that this coupling scheme
allows for highly  efficient quantum state engineering. In
Section~\ref{IC}, we propose the architecture for a distributed
quantum register of many individually trapped ions interconnected
by a cavity field mode. This proposal allows for vibrational quantum
state transfer between two remote ions. We quantitatively address
a non-trivial example. Section~\ref{QBD} is devoted to the
description of a scheme for almost complete Bell-state
measurements performed combining vibrational-mode manipulations and
electronic-state detection. The ability to achieve a
high-efficiency discrimination of the four coherent Bell states
is exploited. In Section~\ref{twoqubits}, we describe how to realize an
entangling two-qubit gate that, together with the single-qubit
rotations, allow for universal coherent quantum computation.

\section{Hamiltonian for quantum state engineering}\label{QSE}
The system we consider is a two-level ion coupled to a
bichromatic field, detuned from the atomic transition. The trap
tightly confines the ion in the $x-y$ plane (as sketched in
Fig.~\ref{systemlevels} {{\bf (a)}). The energy scheme is shown
in Fig.\ref{systemlevels} {{\bf (b)}}. The external fields
(treated classically) illuminate the ion in opposite directions
in the $z=0$ plane and both can have a component along the $x$
and $y$ axes. We assume the trap to be anisotropic, with
$\omega_{x}>\omega_{y}$ and $\Delta{x}_{cm}$ ($\Delta{y}_{cm}$) the ground-state width, in the trapping potential, along the $x$ ($y$) direction. The Hamiltonian of the system reads
($\hbar=1$ is taken throughout this paper)
\begin{equation}
H=\sum_{i=x,y}\omega_{i}\hat{b}^{\dagger}_{i}\hat{b}_{i}+\omega_{eg}
\hat{\sigma}_{+}\hat{\sigma}_{-}+\sum^{2}_{i=1}\left(\hat{g}_{i}\hat{\sigma}_{+}+\hat{g}^{\dagger}_{i}\hat{\sigma}_{-}\right).
\end{equation}
Here, $\hat{b}^{\dag}_{j}$ $(\hat{b}_{j})$ $(j=x,y)$ are the
creation (annihilation) operators describing the quantized
position of the CM of the ion in the trap,
$\hat{g}_{i}=g_{i}{e}^{-i{\bf k}_{i}\cdot{\hat{\bf{ r}}}-i\omega_{i}t-i\phi_{i}}$
take account of the couplings between the ion and the i-th laser ($i=1,2$) of its frequency $\omega_{i}$, wave-vector ${\bf{k}}_{i}\equiv(k_{ix},k_{iy},0)$ and phase $\phi_{i}$. Here, $\hat{\bf r}$ is the vectorial operator of the CM position and $\hat{\sigma}_{+}=\hat{\sigma}^{\dagger}_{-}=\pro{e}{g}$. 
The ion's transition frequency is labelled by $\omega_{eg}$. In a rotating frame 
and in
the limit of large detuning
$\Delta_{1}\gg{g}_{1,2},\delta_{12},\gamma_{rad}$, where
$\Delta_{1}=\omega_{eg}-\omega_{1}$,
$\delta_{12}=\omega_{2}-\omega_{1}$ and $\gamma_{rad}$ the
spontaneous decay-rate of the ion from $\ket{e}$, the atomic
excited state can be adiabatically eliminated.
\begin{figure}[ht]
{\bf (a)}\hskip4.0cm{\bf (b)}
\centerline{\psfig{figure=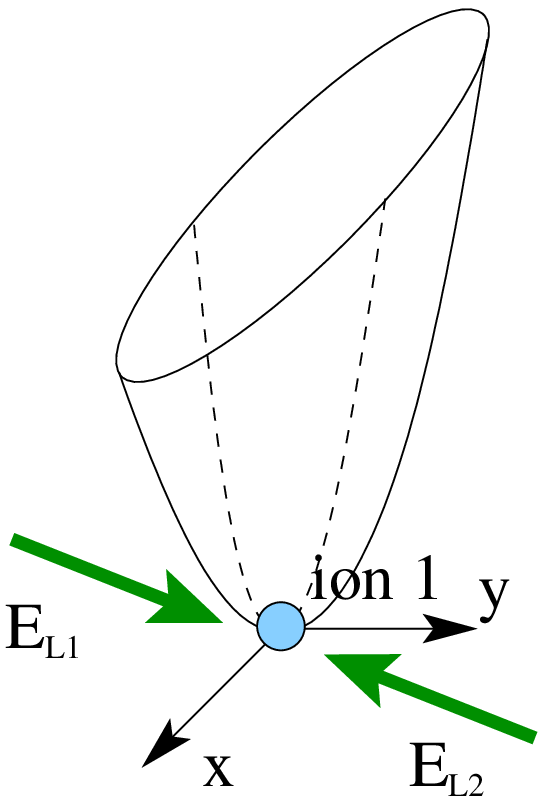,width=2.7cm,height=3.8cm}\hskip2.0cm\psfig{figure=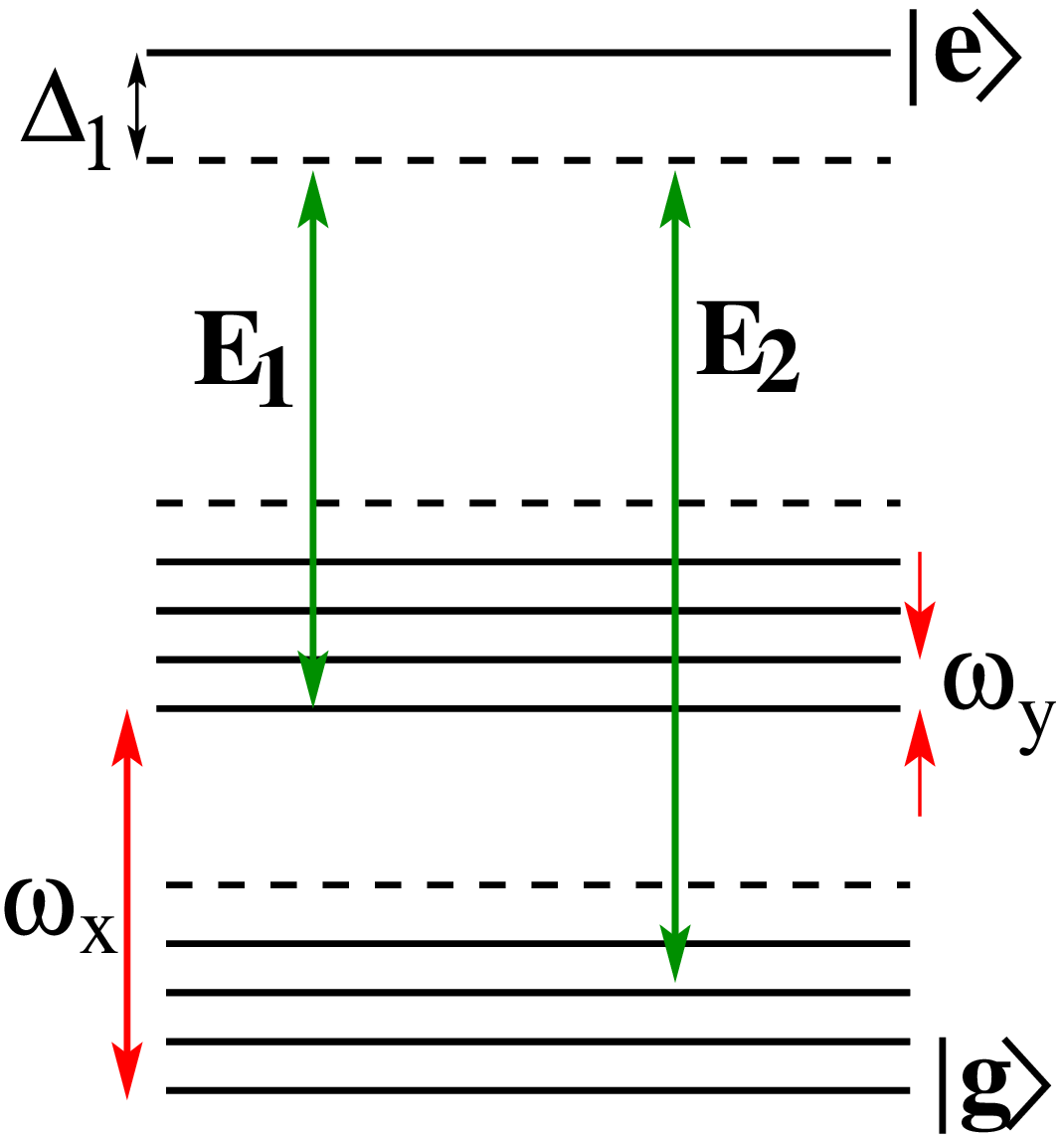,width=2.8cm,height=3.30cm}}
\caption {{\bf (a)}: The physical set-up. A two-level
ion is trapped in a bidimensional trap
($\omega_{y}<\omega_{x}$). Two monochromatic laser
field (classical) excite the ion. {\bf (b)}: The energy level
scheme. The trap is quasi-harmonic so that, in the
limit of resolved sidebands, the different excitations
of the $x$ and $y$ vibrational degrees of freedom are
spaced as shown. The lasers are off-resonance.}
\label{systemlevels}
\end{figure}
After some lengthy calculations and using the Campbell-Baker-Haussdorff
theorem, the Hamiltonian can be written as
\begin{equation}
\label{eq2}
\begin{split}
H&\simeq\sum_{j=x,y}\omega_{j}\hat{b}^{\dag}_{j}\hat{b}_{j}
-\frac{g_{1}g_{2}}{\Delta_{1}}
\sum^{\infty}_{n,m=0}\sum^{\infty}_{p,q=0}\frac{(i{\eta}^\prime_{x})^{n+m}}{n!m!}\\
&\times\frac{(i\eta'_{y})^{p+q}}{p!q!}\hat{b}^{\dag{n}}_{x}
\hat{b}^{m}_{x}\hat{b}^{\dag{p}}_{y}\hat{b}^{q}_{y}e^{-i\delta_{12}t-i\phi}+h.c.
\end{split}
\end{equation}
where $\phi=\phi_{1}-\phi_{2}$ and a numerical factor arising
from the normal ordering has been absorbed in the Rabi
frequencies $g_{1,2}$. Here, $\eta'_{x}=\Delta{x_{cm}}\Delta{k}_{x}$ and $\eta'_{y}=\Delta{y_{cm}}\Delta{k}_{y}$ are 
the effective Lamb-Dicke parameters for the $x$ ($y$) motion respectively~\cite{wineland} and $\Delta{k}_{x,y}$ are the projections of ${\bf k}_{1}-{\bf k}_{2}$ in the $z=0$ plane. We have
neglected the laser-intensity dependent ac-Stark shifts due to
the dispersive couplings. These energy terms in the Hamiltonian
can be controlled by stabilizing the laser beams and formally
eliminated by redefining the ground state energy. A scheme to cancel
the ac-Stark shifts using an additional laser has been
demonstrated in ref.~\cite{schmidtkalerstark}.
 Properly directing the laser beams we can arrange a
coupling between the two vibrational modes as well as engineering a
single-mode Hamiltonian (when $\mu$ or $\nu$ is zero)~\cite{STK}.
In this latter case, if not explicitly specified, we will always
consider the states of the $x$ mode to embody the qubits, while
the $y$ mode will be used as an ancilla. An interesting feature
of the model in Eq.~(\ref{eq2}) is the possibility to select
stationary terms from the Hamiltonian simply by tuning the laser
fields to an appropriate sideband resonance of the trapped ion's
spectrum. Indeed, in the interaction picture, the term depending
on
$\exp\left[i(s_{x}\omega_{x}+s_{y}\omega_{y}-\delta_{12})t\right]$
(and its hermitian conjugate) appears in $H$, where
$s_{x}=(n-m),\,s_{y}=(p-q)$. Tuning $\delta_{12}$, which excites
the proper sideband of the energy-level scheme shown in
Fig.~\ref{systemlevels} {{\bf (b)}}, we single out stationary
terms in Eq.~(\ref{eq2}), which we want to be dominant over the
contributions of the other oscillating terms.

A remarkable range of evolutions is covered by this coupling
scheme and some of them are particularly relevant for the purpose
of coherent quantum computation. We note that, in the protocol
proposed in \cite{jacobmyung}, the leading ingredients are
represented by the ability to generate coherent states and their
macroscopic superpositions ({\it Schr\"odinger cat states}) as
well as multi-mode entangled coherent states. To manipulate the
states of the elements of a quantum register, on the other hand,
ref.~\cite{jacobmyung} prescribes  the use of reliable beam
splitter operations, phase shifts and displacement operations
(these latter effectively perform rotations in the computational
basis). A beam splitter (BS) operation has been described in
ref.~\cite{STK} and we need to give details about 
the engineering of the other operations with our model.

We need a reliable way to generate a coherent state of motion in
order to work in a computational space spanned by the coherent
states $\left\{\ket{\alpha},\ket{-\alpha}\right\}$ (which are
quasi-orthogonal for sufficiently large $\alpha\in\mathbb{C}$).
Many different ways to achieve this have been
suggested~\cite{wineland}. Here, we note that, if
$\delta_{12}=\omega_{x}$ ({\it i.e.} if we excite the first red
sideband of the $x$ motion) and the two fields have
no projection onto the $y$ axis, the stationary term
$H_{cs}\simeq\left(i{g}_{1}g_{2}\eta^{\prime}_{x}/{\Delta_{1}}\right)\hat{b}^{\dag}_{x}+h.c$
is selected, assuming $\phi=0$. This energy term gives rise
to a unitary evolution that corresponds to a displacement
$\hat{D}_{x}(\alpha)=\exp(\alpha\hat{b}_x^\dag-\alpha^*\hat{b}_x)$
in phase-space~\cite{jacobmyung,cochrane}
$\alpha=g_{1}g_{2}\eta^{\prime}_{x}t/\Delta_{1}$ and $t$ the interaction time.
However, if we want to give an estimate of the accuracy of this
{\it state engineering procedure}, the effect of the
non-stationary terms in the Hamiltonian has to be quantitatively
addressed. In order to do it, we consider the
formal relationship between our coupling scheme and the system in
ref.~\cite{STK}, where the three-level $\Lambda$ configuration
can be mapped onto our own coupling-scheme when the
adiabatic elimination of the excited state of the ion is
performed. It is shown in \cite{STK} that by considering an anisotropic trap
with a sufficiently large ratio $\omega_{x}/\omega_{y}$, allows
us to neglect additional {\it accidental} resonances in
Eq.~(\ref{eq2}). To generate a coherent state, we estimate
$\omega_{x}\ge3\omega_{y}$ is enough. On the other hand, the
coupling factors relative to the non-stationary terms in
Eq.~(\ref{eq2}) are sensibly smaller than the rate at which the
coherent state is generated once we guarantee
$(g/\eta_{x}\Delta_{1})\ll\gamma$ with $\gamma=\omega_{x}/g$ a
dimensionless parameter which, experimentally, can be
$\gamma\geq{20}$. For the sake of definiteness, we have assumed
$g_{1}=g_{2}=g$ and, given that $\delta_{12}\ll\omega_{1,2}$, we have taken $\modul{k_{1}}\simeq\modul{k_{2}}=k$ (so that $\eta^{\prime}_{x}\simeq{2\eta_{x}}$). For the realistic value $\eta_{x}=k\Delta{x}_{cm}=0.4$ , the above constraints require $\Delta_{1}\gg{g}/5$.

For our quantitative analysis, we choose $\omega_{x}=4\omega_{y}$
and $\Delta_{1}=5{g}$.
Then, we retain the terms, in $H$, which oscillate at the
frequencies $\omega_{x},\,\omega_{y},\,\omega_{x}{\pm}\omega_{y}$
so that the dynamic generator we consider is
$H_{true}=H_{cs}+H_{nst}$, where $H_{nst}$ collects the
non-stationary terms we discussed.  With the above choices for
the relevant parameters, we look for the overlap
$\modul{\sprod{\alpha}{\psi}_{t}}$ between the coherent state
$\ket{\alpha}$ we want to generate and the state
$\ket{\psi}_{t}=\exp\left\{-iH_{true}t\right\}\ket{0}_{x}$.
Numerically, we are limited by the dimension of the computational
space. We thus take $\modul{\alpha}=1$ and truncate
the basis to $\left\{\ket{0},..,\ket{5}\right\}_{x}$, with
$\ket{n}_{x}$ ($n=0,..,5$) indicating phonon-number states. On
the other hand, the state of the $y$ mode should not be affected
at all by the desired evolution. Assuming $\ket{0}_{y}$ for the
initial state, it is reasonable to take
$\left\{\ket{0},\ket{1}\right\}_{y}$ for
the evolution of this state due to $H_{true}$. The motional state
of the ion at time $t$ is therefore expanded as
$\ket{\psi}_{t}=\sum^{5}_{n,0}\sum^{1}_{m,0}A_{nm}(t)\ket{n,m}_{xy}$,
with $\sum^{5}_{n,0}\sum^{1}_{m,0}\modul{A_{nm}}^{2}=1$. The
overlap reads ${\cal O}(t)={\cal
N}\sum^{5}_{n=0}A_{n0}(t)/\sqrt{n!}$, (with ${\cal N}$ a
normalization factor) that can be evaluated once we solve the set
of differential equations obtained projecting the Schr\"odinger
equation for $\ket{\psi}_{t}$ onto the $\ket{n,m}_{xy}$ states.
The result is shown in Fig.~\ref{overlap2} {\bf{(a)}}. The overlap
becomes perfect when the rescaled interaction time is
$gt=\Delta_{1}/(g\eta^{\prime}_{x})=6.25$. Furthermore, we have checked
that with the above values the efficiency of the process is
insensitive to an increase of the Lamb-Dicke parameter up to
$\eta_{x}\simeq{0.9}$. For a larger $\eta_{x}$, some deviation
from the ideal case is observed. Another parameter that is
relevant in this investigation is $\gamma$. Reducing it means
lowering the oscillation frequencies of the non-stationary terms
in $H_{true}$.  This spoils the efficiency of the
entire state-engineering  process. An example of this effect is
given in Fig.~\ref{overlap2} {\bf{(b)}}. It is worth stressing 
that, even though the generation of a coherent state with just 
a small amplitude $\alpha$ has been checked here, this will also 
apply for a larger (in principle arbitrary) amplitude.
\begin{figure}
{\bf{(a)}}\hskip4.7cm{\bf{(b)}}
\centerline{\psfig{figure=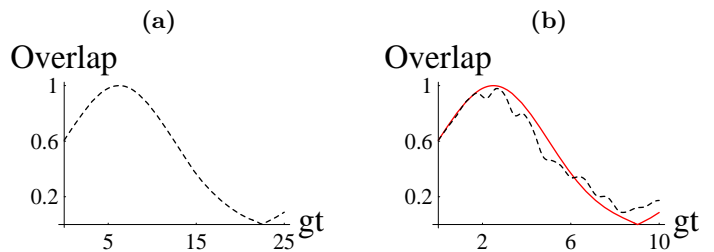,width=10cm,height=3.0cm}}
\caption {{\bf (a)}: The overlap between a coherent state
$\ket{\alpha=1}_{x}$ and the state generated by the non-ideal
Hamiltonian $H_{true}$ that contains non-stationary terms at low
frequencies up to second order in the Lamb-Dicke parameter.
In this simulation, $\eta_{x}=0.4$. Other parameters are as in
the text. {\bf (b)}: Comparison between the perfect overlap
(solid line) and the one obtained with $\eta_{x}=1.0$ and
$\gamma=10$ (dashed line).} \label{overlap2}
\end{figure}

For a single qubit operation, we first consider the rotation
$U^{z}(\theta/2)$ around the $z$ axis of the Bloch sphere for the qubit
$\left\{\ket{\pm\alpha}\right\}_{x}$.  This rotation 
is very well approximated by displacement operation
$\hat{D}_{x}(i\epsilon)$, with $\epsilon=\theta/4\alpha$
$\left(\theta\in{[}0,2\pi]\right)$~\cite{jacobmyung}.
\begin{equation}
\label{rot}
U^{z}\left(\frac{\theta}{2}\right)=
\begin{pmatrix}
e^{i\frac{\theta}{2}}&0\\
0&e^{-i\frac{\theta}{2}}
\end{pmatrix}.
\end{equation}
Indeed, $\hat{D}(i\epsilon)\ket{\alpha}_{x}=
e^{2i\epsilon\alpha}e^{(\alpha+i\epsilon)\hat{b}^{\dag}_{x}-(\alpha-i\epsilon)
\hat{b}_{x}}\ket{0}_{x}=e^{2i\alpha\epsilon}\ket{\alpha+i\epsilon}_{x}\simeq{e}^{i\frac{\theta}{2}}\ket{\alpha}_{x}$,
where the condition $\alpha\gg\epsilon$ has been assumed
(keeping the product $\alpha\epsilon$ always finite). In
practice, for $\alpha\gtrsim{2}$, a small $\epsilon$ is
sufficient to perform a $2\pi$ rotation.

The rotation $\hat{U}^{x}(\pm\pi/4)$~\cite{jacobmyung} around
 the $x$ axis of the qubit's Bloch sphere can be performed by a
Kerr interaction $H_{K}=\chi_{K}(\hat{b}^{\dag}_{x}\hat{b}_{x})^{2}$. We now
briefly describe the procedure to obtain such a Hamiltonian.
Aligning the laser beams along the $x$ axis (to get
$\mu=1,\,\nu=0$) and arranging their relative detuning
$\delta_{12}=0$, we select a stationary term proportional to
$\hat{b}^{\dagger{2}}_{x}\hat{b}^{2}_{x}$ in Eq.~(\ref{eq2}),
with the corresponding rate of non-linearity
$\chi_{K}=2g_{1}g_{2}\eta^{\prime4}_{x}/\Delta_{1}$. We exploit the
canonical commutation rules between $\hat{b}^{\dag{}}_{x}$ and
$\hat{b}^{}_{x}$ and the relation
$[\hat{n}_{x},\hat{n}^{2}_{x}]=0$
($\hat{n}_{x}=\hat{b}^{\dag}_{x}\hat{b}_{x}$), to obtain
$\exp\left\{-iH_{K}{t}\right\}\vert_{\chi{t}=\pi/2}\ket{\mp{i}\alpha}_{x}=\frac{1}{\sqrt
2}\left(\ket{\alpha}\pm{i}\ket{-\alpha}\right)_{x}$~\cite{yurkestoler}.
This macroscopic superposition of coherent states is the
result of the rotation in the Bloch sphere $\{|\pm\alpha\rangle\}$.

In order to check the effects of the non-resonant terms in the Hamiltonian
of our Kerr non-linear evolution, we have conducted an
analysis similar to the one previously performed to generate 
a coherent state. This time,
$H_{true}=H_{K}+H_{nst}$ contains terms oscillating at the lowest
frequencies and up to the fourth power in $\eta_{x}$.
The stationary term $H_{K}$ dominates over $H_{nst}$ because the
effect of the high-frequency oscillating terms is
averaged out from the effective dynamical evolution of the qubit.
We retain the same values used before for the relevant parameters
in our calculations, showing that they are suitable for this
effective rotation too. Our model is sufficiently flexible not to
require further adjustments of the set-up. We consider the
transformation
$\ket{\alpha=-i}_{x}\rightarrow\ket{cat1}_{x}=\frac{1}{\sqrt
2}\left(\ket{\alpha=1}+i\ket{\alpha=-1}\right)_{x}$ and use  the
truncated phonon-number basis
$\left\{\ket{0},..,\ket{5}\right\}^{\otimes{2}}$ to solve the
projected Schr\"odinger equations
\begin{equation}
i\,_{xy}\!\bra{p,q}\partial_{t}\ket{\psi(t)}_{xy}=\,_{xy}\!\bra{p,q}H_{true}\ket{\psi(t)}_{xy},
\end{equation}
with $\ket{\psi(0)}_{xy}=\ket{-i}_{x}\ket{0}_{y}$ and the
decomposition
$\ket{\psi(t)}_{xy}=\sum^{5}_{n,m=0}A_{nm}(t)\ket{nm}_{xy}$. The
normalization of the wave-function implies
$\sum^{5}_{n,m=0}\modul{A_{nm}(t)}^{2}=1$, as before.
\begin{figure}
\centerline{\psfig{figure=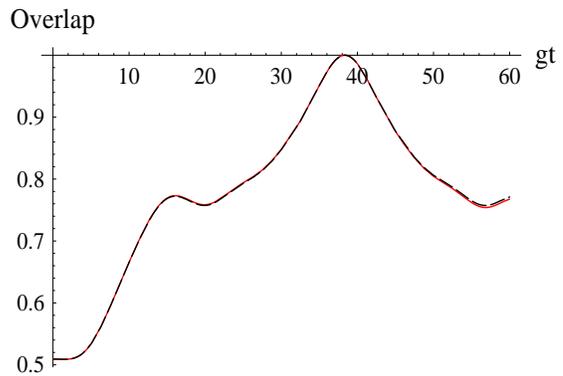,width=7.5cm,height=5.0cm}}
\caption {Comparison between the perfect overlap ${\cal O}_{id}(t)$
(broken line) and the true one ${\cal O}_{true}(t)$ obtained
numerically solving the Schr\"odinger equation governed by
$H_{true}$. In these calculations, $\eta_{x}=0.4$ and $g=500\pi\,KHz$. Other parameters are as
described in the body of the paper.} \label{overlapcat}
\end{figure}
In Fig.~\ref{overlapcat} we compare the ideal overlap ${\cal
O}_{id}(t)=\modul{_{x}\!\sand{cat1}{e^{-iH_{K}t}}{-i}_{x}}$
(dashed line) to the overlap  ${\cal
O}_{true}(t)=\modul{_{xy}\!\bra{cat1,0}{e^{-iH_{true}t}}\ket{-i,0}_{xy}}$
(solid curve), whose time behavior depends on the coefficients
$A_{nm}(t)$, for counter-propagating lasers with $g_{1,2}=g$ and $\eta_{x}=0.4$. It is apparent
that the dynamical evolution governed by $H_{true}$ leads to the
desired superposition state at the expected rescaled time
$gt=5\pi/2\eta^{\prime{4}}_{x}\simeq{38}$. The match between the two
curves is very good up to the rescaled interaction times we show.
Being a rotation in the space spanned by the coherent states with
amplitude $\alpha=\pm{1}$,
the behavior of the overlap is periodic and replicates itself for
interaction times larger than those shown in
Fig.~\ref{overlapcat}. This numerical simulation suggests that
Kerr non-linear interactions and, thus, the
$U^{x}(\pm\pi/4)$ rotation, can actually be performed quite
efficiently in this set-up. We estimate
$\chi_{K}
\simeq20\pi\,KHz$, for
$g=500\pi\,KHz$~\cite{wineland,sasura}, corresponding to an
interaction time $t=\pi/2\chi_{K}\simeq25\,\mu{sec}$. On the
other hand, the effective life-time of the excited state
$\ket{e}$, in the non-resonant regime we are considering, is
$\tau_{spont}={\Delta^{2}_{1}}/({{g}^2}\gamma_{rad})\simeq{1}\,sec$
if we use metastable levels of an optical transition (for example
the $S_{1/2}\rightarrow{D}_{5/2}$ transition in $^{40}Ca^{+}$
whose excited level has a natural life-time of about $1\,sec$
($\gamma_{rad}\simeq{1\,sec}$)~\cite{wineland,schmidt-kaler,schmidtkalerstark}).

We need here to make a remark: the analysis we have
performed always assumes that the ancillary $y$ mode is prepared
in the vacuum state. This is just for mathematical convenience.
We have derived the equations of motion for the case of an initial  
coherent state $\ket{\beta=1}_{y}$ of the $y$-motional mode. Here,
again, a small amplitude of the coherent state is taken because
it is then possible to truncate the computational  phonon-number
basis, considerably simplifying the calculations. We have
concluded that the comparison between $Ov_{id}(t)$ and
$Ov_{true}(t)$ shows the same qualitative features seen in
Fig.~\ref{overlapcat}. We conjecture that the same conclusion
holds regardless of the state in which the $y$ degree of motion
has been prepared, if the parameter in $H_{true}$ are kept within
the range of validity of the approximations above.

With arbitrary rotations around the $z$ axis (implemented via
effective displacements) and $\pi/4$ rotations around the $x$
axis, it is actually possible to build up any desired rotation
around the $y$ axis of the Bloch sphere. This, in turn, allows us
to arbitrarily rotate the qubit around the $x$
axis~\cite{jacobmyung}. The two operations we have demonstrated
are thus sufficient to realize any desired one-qubit
rotation. In particular, the sequence
$U^{z}(\pi/4)U^{x}(\pi/4)U^{z}(\pi/4)\ket{\pm\alpha}_x$ realizes
the transformation $\ket{\pm\alpha}_{x}\!\rightarrow(1/\sqrt{2})
\left(\ket{\alpha}_x\pm\ket{-\alpha}_x\right)$ that is a 
Hadamard gate. The resulting states, here, are the the so-called 
{\it even (for $+$ sign) and odd (for $-$ sign) coherent states} 
as they are the superposition of just
even and odd phonon-number states, respectively~\cite{ioEIT}. An
interesting feature that will be exploited later is that even and
odd coherent states are eigenstates of the parity operator
$(-1)^{\hat{n}}$ ($\hat{n}$ is the phonon-number operator) with
eigenvalue $\pm{1}$, respectively. We will discuss later the role 
these states have in coherent quantum computation.

As a final relevant case treated here, we now consider the
engineering of a  bimodal non-linear interaction suitable for the
generation of entangled coherent states (ECS)~\cite{sanders}.
This class of states will be represented as
\begin{equation}
\label{canoniciecs}
\begin{split}
\ket{\phi_{\pm}}=N^{\phi}_{\pm}\left\{\ket{\alpha,\alpha}\pm\ket{-\alpha,-\alpha}\right\},\\
\ket{\psi_{\pm}}=N^{\psi}_{\pm}\left\{\ket{\alpha,-\alpha}\pm\ket{-\alpha,\alpha}\right\}.\\
\end{split}
\end{equation}
States $\ket{\phi_{\pm}}$ and $\ket{\psi_{\pm}}$ can be generated
by  superimposing, at a $50:50$ BS, a zero-phonon
state $\ket{0}$ with an even and odd coherent state,
respectively. As an example, suppose that, via the procedure
described above, we have created an even coherent state of the
$x$ motional mode while $y$ is in its vacuum state. Arranging a
BS interaction between $x$ and $y$ phonon
modes~\cite{STK}, the joint state
of the two vibrational modes is then transformed into one of 
the entangled coherent states above. Alternatively, $\ket{\phi_{\pm}}$ 
and $\ket{\psi_{\pm}}$
can be produced using the cross-phase modulation Hamiltonian
$H_{cp}=\chi_{cp}\hat{n}_{x}\hat{n}_{y}$, with $\chi_{cp}$
the rate of non-linearity~\cite{ioEIT}. Starting from
$\ket{\alpha}_{x}\ket{\beta}_{y}$, this interaction produces
$\ket{ecs}_{xy}\propto\ket{\alpha}_{x}\left(\ket{\beta}+\ket{-\beta}\right)_{y}+
\ket{-\alpha}_{x}\left(\ket{\beta}-\ket{-\beta}\right)_{y}$ when
$\chi_{cp}t=\pi$~\cite{ioEIT}, which can be reduced to the form of ECS in
Eq.~(\ref{canoniciecs}) through single-qubit manipulation. Thus,
having already discussed how to perform one-qubit operations,
we concentrate here on the generation of this kind of generalized
ECS.

By inspection of $H_{cp}$, we recognize the necessity of an 
interaction symmetry in the two vibrational modes. 
This can be obtained directing the lasers
 at $45$ and $225$ degrees with respect to the $x$ axis. In this case,
$\delta_{12}=0$ has to be set in order to select the
stationary term
$\chi_{cp}\hat{b}^{\dag}_{x}\hat{b}_{x}\hat{b}^{\dag}_{y}\hat{b}_{y}$
in Eq.~(\ref{eq2}) with
$\chi_{cp}=2g_{1}g_{2}\eta^{\prime 2}_{x}\eta^{\prime 2}_{y}/\Delta_{1}$. The
other terms in the Hamiltonian are rapidly oscillating and
negligible if the same dynamical conditions we commented above
are assumed. We consider ${\cal
O}_{ecs}(t)=\modul{_{xy}\!\bra{ecs_{id}}{e^{-iH_{true}t}}\ket{\alpha=1,\beta=1}_{xy}}$,
with $H_{true}=H_{cp}+H_{nst}$ the Hamiltonian containing both
the desired non-linear interaction $H_{cp}$ and all the relevant
non stationary terms. We have taken
$\ket{ecs_{id}}_{xy}\!\propto\!\ket{1}_{x}\left(\ket{1}+\ket{-1}\right)_{y}+\ket{-1}_{x}\left(\ket{1}-\ket{-1}\right)_{y}$,
where all the states appearing in this expression are coherent states
of their amplitude $\modul{\alpha}=\modul{\beta}=1$. To evaluate
${\cal O}_{ecs}(t)$, the computational basis has been truncated to
$\left\{\ket{0},..,\ket{5}\right\}^{\otimes{2}}$, as usual.
The results are shown in Fig.~\ref{overlapecs}. The dashed line represents the
ideal behavior of the overlap, that is, its time dependency when
just the ideal interaction $H_{cp}$ is considered. This curve is
contrasted with the overlap obtained when the full Hamiltonian
$H_{true}$ is taken. The mismatches between the curves  are very
small and the overall comparison is excellent. Here,
$\eta_{x}=\eta_{y}=0.4$ and all the other relevant parameters are
the same as in the previous simulations. The rescaled time
$gt\simeq77$, where the overlap is almost perfect, is
equivalent to an effective interaction time of
$t\simeq{50}\,\mu{sec}$, using the same parameters of the previous
calculations. The scheme appears, thus, to be robust against the
spoiling effects of the non-stationary terms and is efficient
within the coherence times of the physical system we consider.
\begin{figure}[ht]
\centerline{\psfig{figure=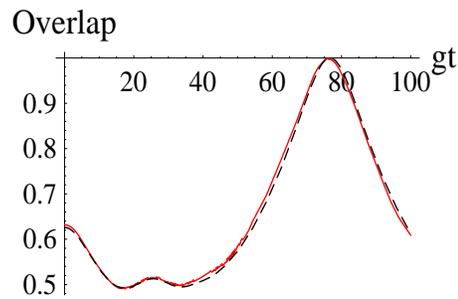,width=6.0cm,height=4.0cm}}
\caption {Comparison between the perfect overlap  (dashed line)
and the true one ${\cal O}_{ecs}(t)$ obtained solving the
Schr\"odinger equation governed by $H_{true}$. In this
simulation, $\eta_{x,y}=0.4$ and $g_{1,2}=g=500\pi\,{KHz}$. Other parameters as described in the body
of the paper.} \label{overlapecs}
\end{figure}
\section{Coupling between motional degrees of freedom of individually trapped ions}
\label{IC} So far, our discussions have been limited to the case
of a single ion. Unfortunately, considering the vibrational modes of
just a single trapped ion is a substantial limitation on the
computational capabilities of our device. However, we can take an
advantage  of some recent experiments that demonstrate the
coupling between trapped ions and a high-finesse optical
cavity~\cite{walther,schmidt-kaler} to design the register of a
vibrational quantum computer as formed by several remote and 
independently trapped ions. Here, we describe in detail a
mechanism to couple the motional degrees of freedom of different
elements of such a quantum register. In the experiments reported
in refs.~\cite{walther,schmidt-kaler}, a coherent interaction is
established between an ion and a cavity mode. The optical
transition between metastable states of a ${Ca}^{+}$ ion has been
used to embody a qubit and the atomic responses to both temporal
and spatial variations of the coupling have been
analyzed~\cite{schmidt-kaler}. These impressive experimental
achievements justify and motivate {\it a posteriori} the
two-level model in the optical range of frequency we have
assumed here. The system sketched in
Fig.~\ref{trappolacavity}, is based on the linear geometry of the
ion trap-optical cavity interfaces demonstrated
in~\cite{walther,schmidt-kaler}. This set-up is suitable to store 
a linear {\it ion crystal}
represented by a row of aligned traps, which are mutually independent and
spatially well-separated. The cavity field mode (here assumed to
be a transverse $\mbox{TEM}_{00}$ mode of a near-confocal
resonator) is described by its bosonic annihilation (creation)
operator $\hat{a}$ ($\hat{a}^{\dag}$) and is
aligned with the $x$-axis of the bidimensional traps. The
interaction with each ionic transition is off-resonant with
detuning $\Delta_{i}$ ($i=1,2$) respectively (assumed to be different for 
sake of generality. The mathematical approach is simplified if $\Delta_{1}=\Delta_{2}$). Two external fields, $E_{L1},\,E_{L2}$ excite the ions and are directed along
the $y$-axis. As we will see, this effectively couples the $y$
modes of the ions.

We assume a standing-wave configuration for the spatial
distribution of the cavity field with the ions placed at the
nodes of a cosine function~\cite{commento,sasura}. In a rotating frame
at the frequency $\omega_{L1}$ of the laser $E_{L1}$ and in the
interaction picture with respect to the free energy
 of the resonator, the
Hamiltonian of our system reads
\begin{equation}
\label{HamCav}
\begin{split}
H_{ic}=&\sum_{j=x,y}\sum^{2}_{i=1}\omega_{ji}\hat{b}^{\dag}_{ji}
\hat{b}_{ji}-\frac{{g}_{0}{\cal
E}_{L1}\eta_{y1}}{\Delta_{1}}\left(\hat{b}^{\dag}_{y1}+\hat{b}^{}_{y1}\right)\hat{a}^{\dag}e^{i\delta_{c}t}\\&-\frac{{g}_{0}{\cal
E}_{L2}\eta_{y2}}{\Delta_{2}}\left(\hat{b}^{\dag}_{y2}+\hat{b}^{}_{y2}\right)\hat{a}^{\dag}e^{i(\delta_{c}+\Delta_{12})t+i\phi}+h.c.
\end{split}
\end{equation}
Here, $\delta_{c}=\omega_{cavity}-\omega_{L1}$ and $\Delta_{12}=\omega_{L1}-
\omega_{L2}$ while ${\cal E}_{Li}\,(i=1,2)$ is the Rabi
frequency of the interaction between ion $i$ and laser $E_{Li}$
and $\phi$ is the phase-difference between the lasers. The condition
$\Delta_{1,2}>{\cal E}_{L1},{\cal E}_{L2},g_{0}$ underlies Eqs.~(\ref{HamCav}), where the electronic excited states of the ions have been adiabatically
eliminated. An intuitive picture of the dynamics of the system in
this set-up is gained in the overdamped-cavity regime (or {\it
bad cavity limit}), where the cavity decay-rate $\kappa$ is
much larger than any other rate involved in Eq.~(\ref{HamCav}).
\begin{figure}
\centerline{\psfig{figure=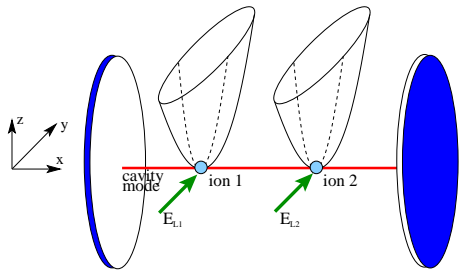,width=6.0cm,height=3.5cm}}
\caption {The set-up for a motional state transfer procedure.
Two trapped ions are placed along the $x$ axis of a mono-mode
optical cavity having a standing-wave configuration. The ions are
coupled to the cavity field mode and to two external laser
pulses. The laser fields illuminate the ions along the $y$ direction.
This realizes an effective coupling between the $y$ vibrational
modes of the trapped ions.} \label{trappolacavity}
\end{figure}
In this case, the cavity mode, which is detuned from the ionic
transitions, represents an off-resonant bus that is only
virtually excited by the interactions with the ions and can be
eliminated from the dynamic of the overall system. In a formal
way, we can consider the evolution equation of the cavity field
operator $\hat{a}^{\dag}$ and impose that its variations are
negligible within the time-scale set by the effective coupling
$\Omega_{i}=g_{0}{\cal E}_{Li}\eta_{yi}/\Delta_{i}$ ( that implies the cavity
field mode has already reached its stationary state). This results in an effective interaction that, in a
rotating frame at the frequencies of the traps (supposed to be
identical for ion 1 and 2), reads
\begin{equation}
\label{HamCavRWA}
H^{rwa}_{ic}\simeq-\sum^{2}_{i\neq{j},1}
\frac{{g}^2_{0}{\cal E}_{Li}{\cal
E}_{Lj}\eta_{yi}\eta_{yj}}{\Delta_{i}\Delta_{j}\kappa}\left(\hat{b}^{\dag}_{yi}\hat{b}_{yj}e^{i\phi}
+\hat{b}^{\dag}_{yj}\hat{b}^{}_{yi}e^{-i\phi}\right),
\end{equation}
where the condition
$\Delta_{12}=\omega_{y2}-\omega_{y1}$ and the Rotating Wave
Approximation (RWA) have been used. This interaction models a
BS operation between motional degrees of freedom
belonging to spatially separated  trapped ions. This interaction
is useful for entanglement generation and motional state
transfer, where the states $\ket{\psi}_{y1}$ and $\ket{0}_{y2}$
are swapped, with $\ket{\psi}_{y1}$ being completely arbitrary.
This is exactly what we want to realize for the purpose of motional state
transfer. If the ion crystal is larger than two units, two
specific ions can be connected by exciting them (and only them) with
the laser fields. The other trapped ions will be unaffected by
the coupling. Once the local interaction between the $x$ and $y$
motional modes of a {\it specific} ion has been performed
(according to a given quantum computing protocol), then the state
of the $y$ mode can be properly transferred to another ion of the crystal,
labelled $l$, that has been prepared in $\ket{0}_{yl}$.
However, for the sake of realism, in what follows we
pursue the analysis restricted to a two-ion system and give some
more insight in the process of motional state transfer.

The assumed bad cavity limit is particularly convenient to
isolate the dynamics of the motional modes from that of the bus.
Indeed, a full picture of the evolution of the system is gained by the master equation (in the interaction picture)
\begin{equation}
\label{ME}
\partial_{t}\rho=-i\left[H'_{ic},\rho\right]+\kappa\left(2\hat{a}\rho\hat{a}^{\dag}
-\left\{\hat{a}^{\dag}\hat{a},\rho\right\}\right)=(\hat{\cal
L}_{0}+\hat{\cal L}_{cav})\rho,
\end{equation}
with $\rho$ the total density matrix of the
$ion\,1+ion\,2+cavity$ system and, taking $\delta_{c}=\omega_{y1},\,\Delta_{12}=\omega_{y2}-\omega_{y1}$, it is $H'_{ic}=-\sum^{2}_{i=1}\Omega_{i}(\hat{b}_{yi}\hat{a}^{\dag}+h.c.)$. We have used the notation $\hat{\cal L}_{0}\rho=-i[H'_{ic},\rho]$. We now go to a dissipative picture defined by
$\tilde\rho=e^{-\hat{\cal L}_{cav}t}\rho$ and exploit the
relations $\hat{\cal L}_{cav}[\hat{a},\rho]=[\hat{a},(\hat{\cal
L}_{cav}-\kappa)\rho]$, $\hat{\cal
L}_{cav}(\hat{a}\rho)=\hat{a}\hat{\cal
L}_{cav}\rho+\kappa\hat{a}\rho$ (and analogous for
$\hat{a}^{\dag}$)~\cite{iotransfer}. After some lengthy
calculations, Eq.~(\ref{ME}) reduces to
\begin{equation}
\label{MEC}
\begin{split}
\partial_{t}\tilde\rho&=i{g}_{0}\sum^{2}_{i=1}\frac{{\cal E}_{Li}
\eta_{yi}}{\Delta_{i}}\left\{e^{-\kappa{t}}[\hat{b}_{yi},\tilde\rho]
\hat{a}^{\dag}+e^{\kappa{t}}\hat{b}_{yi}[\hat{a}^{\dag},\tilde\rho]-
h.c.\right\}\\&\equiv{e}^{-\kappa{t}}\hat{\cal L}_{1}\tilde\rho+e^{\kappa{t}}\hat{\cal L}_{2}\tilde\rho,
\end{split}
\end{equation}
with $\hat{\cal L}_{1}$ ($\hat{\cal L}_{2}$) an effective
super-operator obtained by collecting all the terms in
Eq.~(\ref{MEC}) having the $e^{-\kappa{t}}$ ($e^{\kappa{t}}$)
pre-factor. To isolate the vibrational degrees of freedom, we trace
over the cavity mode. We obtain $\partial_{t}\rho_{v}=\hat{\cal
L}_{1}(e^{-\kappa{t}}\tilde\rho)$, with
$\rho_{v}=Tr_{cav}(\tilde\rho)$.
This master equation still involves the cavity variables because
of the presence of $\tilde\rho$. In order to remove these dependencies, we
go back to Eq.~(\ref{ME}),  integrate it formally and multiply it
by $e^{-\kappa{t}}$. In the limit of large $\kappa$, we can
invoke the first Born-Markov approximation and set
$\tilde\rho=\rho_{v}\otimes\rho_{cav,ss}$ with $\rho_{cav,ss}$
the steady state of the cavity mode. It is, then,
$e^{-\kappa{t}}\tilde\rho(t)\simeq\int^{\infty}_{0}\hat{\cal
L}_{2}(\rho_{v}(t)\otimes\rho_{cav,ss})e^{-\kappa{t'}}dt'$ and
\begin{equation}
\label{ME1}
\begin{split}
&\partial_{t}\rho_{v}=Tr_{cav}\left\{\hat{\cal L}_{1}\left(\int^{\infty}_{0}\hat{\cal L}_{2}
(\rho_{v}(t)\otimes\rho_{cav,ss})e^{-\kappa{t'}}dt'\right)\right\}\\
&=\sum^{2}_{i,j=1}\frac{g^{2}_{0}{\cal E}_{Li}{\cal E}_{Lj}\eta_{yi}\eta_{yj}}
{\Delta_{i}\Delta_{j}\kappa}\left\{\left[\hat{b}_{yi},\rho_{v}
\hat{b}^{\dag}_{yj}\right]-\left[\hat{b}^{\dag}_{yi},\hat{b}_{yj}\rho_{v}\right]\right\}\\
&\equiv\sum^{2}_{i=1}\Gamma_{i}\left\{2\hat{b}_{yi}\rho_{v}
\hat{b}^{\dag}_{yi}-\rho_{v}\hat{b}^{\dag}_{yi}\hat{b}_{yi}-\hat{b}^{\dag}_{yi}\hat{b}_{yi}\rho_{v}\right\}+\\
&\sqrt{\Gamma_{1}\Gamma_{2}}\left[2\hat{b}_{y1}\rho_{v}\hat{b}^{\dag}_{y2}+2\hat{b}_{y2}
\rho_{v}\hat{b}^{\dag}_{y1}-\left\{\hat{b}^{\dag}_{y1}\hat{b}_{y2}+\hat{b}^{\dag}_{y2}\hat{b}_{y1},\rho_{v}\right\}\right]
\end{split}
\end{equation}
with $\Gamma_{i}=\Omega^{2}_{i}/\kappa$ ($i=1,2$). This is the
reduced master equation in our study. So far, we have not
included the relaxation terms due to the decay of the motional
amplitude of the ion modes. However, these can be included in the
above derivation simply adding to Eq.~(\ref{ME}) the Liouvillian
terms proportional to the vibrational decay-rate $\gamma_{v}$
(assumed to be equal for the $x$ and $y$ motion). These terms do
not depend  on the cavity operators and are left unaffected by
the adiabatic elimination of the field mode. Their inclusion will
eventually result in a modification of just the {\it single mode
effective rates} according to $\Gamma_{i}\rightarrow\Gamma_{i}+\gamma_{v}$. 
Of course, there is still much more to understand about the mechanisms 
that lead to vibrational decoherence~\cite{murao}. However, some 
estimates of $\gamma_{v}$ put it in the range of tens of milliseconds~(see
\cite{wineland} and Roos {\it et al.} in~\cite{murao}) and, as we
will see, we estimate them to be longer than the effective
interaction times required for a complete motional state
transfer. Thus, from now on, we drop $\gamma_{v}$ from our
analysis. Eq.~(\ref{ME1}) can be projected onto the phonon-number 
basis $\left\{\ket{n,m}\right\}_{y1,y2}$ to give effective
evolution equations that are used for a numerical estimation of
the dynamics of ${\rho}_{v}$.

As an example of motional state transfer, we quantitatively
address the case of
$\ket{\psi}_{y1}=\sqrt{2/5}\left(\ket{0}-\ket{1}\right)_{y1}+ 
\sqrt{1/5}\ket{2}_{y1}$, where
$(|0\rangle,~|1\rangle,~|2\rangle)_{y1}$ are phonon-number states, 
being prepared in ion $1$. Here, the choice has been completely arbitrary
(any other state could have been taken). However, this example
offers us the possibility to see the influence of our protocol on
relative phases in general linear superpositions. Furthermore,
the possible leakage into the Hilbert space complementary to the
one spanned by our computational basis can be investigated. For
quantitative calculations, we restrict the basis to
$\left\{\ket{0},..,\ket{3}\right\}^{\otimes{2}}$. For the
transfer protocol to be effective, the state of the $y2$ mode
must be prepared in $\ket{0}_{y2}$. Some comments are necessary
in order to clarify the protocol. From now on, we refer to the
ion whose $y$ motional state has to be transferred as the {\it
transmitter} while the {\it receiver} is the ion prepared in
$\ket{0}_{y}$. The interaction channel between the transmitter and
the receiver is open if and only if both the ions are illuminated by
the external laser fields. This means that, once one
of the lasers is turned off, the transfer process stops and
the interaction channel is interrupted and unable to further
affect the joint state of the two ions. On the other hand, the
effective interaction has to last for a time sufficient to
complete the transfer.
We find that temporally counter-intuitive laser pulses have to be
applied to the system of transmitter-receiving ions. In
particular, an efficient motional state transfer is achieved if
the effective coupling rate $\Omega_{2}$ decreases while
$\Omega_{1}$ increases in such a way that
$\int_{\mathbf{T}}{dt'}\sqrt{\Gamma_{1}(t')\Gamma_{2}(t')}=\vartheta=\pi/2$,
where $\mathbf{T}$ is the total interaction period and 
time-dependent laser pulses have been assumed. An example of
such pulses is given by
$\Gamma_{1}(t)=\Gamma_{2}(-t)=\tilde\Gamma{e^{-\tilde\Gamma{t}}}/\left({e^{\tilde
\Gamma{t}}+e^{-\tilde\Gamma{t}}}\right)$~\cite{mabuchi}.
We have assumed ${\cal E}_{L1}={\cal E}_{L2}={\cal E}_{L}$,
$\eta_{y1}=\eta_{y2}=\eta_{y}$ and
$\Delta_{1}=\Delta_{2}=\Delta$, so that
$\tilde\Gamma=2g^2_{0}{\cal E}^2_{L}\eta^2_{y}/\Delta^2\kappa$.
The time behaviors of $\Gamma_{1},\,\Gamma_{2}$ and
$\sqrt{\Gamma_{1}\Gamma_{2}}$ are shown in Fig.~\ref{impulsi} {\bf
(a)}. To satisfy the conditions for the adiabatic elimination of
the cavity mode ($\Omega_{1},\,\Omega_{2}\ll{\kappa}$), we have
taken $\tilde\Gamma=0.03$.

\begin{figure} [ht]
\centerline{\psfig{figure=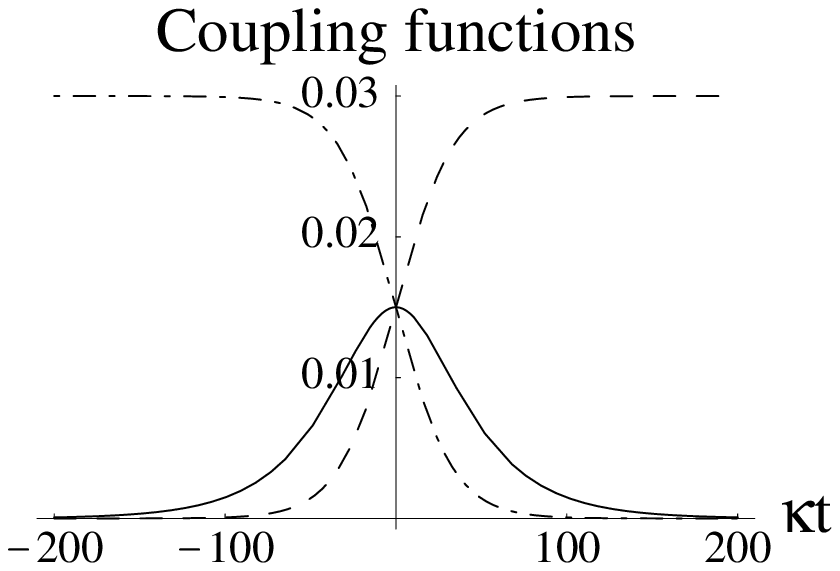,width=4.5cm,height=3.0cm}\psfig{figure=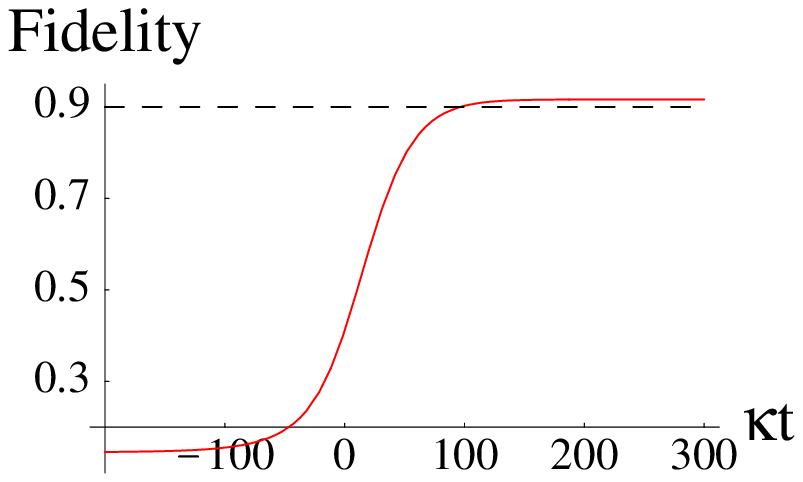,width=4.5cm,height=3.0cm}}
\caption {{\bf (a)}: We plot $\Gamma_{1}(t)$ (dashed line), $\Gamma_{2}(t)$ (dot-dashed) and $\sqrt{\Gamma_{1}(t)\Gamma_{2}
(t)}$ (solid line) as a function of the rescaled interaction time
$\kappa{t}$ and for $\tilde\Gamma=0.03$. The {\it interaction channel}
is open at $t=-200\kappa^{-1}$. The
fidelity does not reach $1$ because of the
effective dissipative part in the evolution of the vibrational state
that, actually, raises the vacuum component of the joint vibrational
state.  {\bf(b)}: Fidelity
of motional state transfer $\ket{\psi,0}_{y1,y2}\rightarrow\ket{0,\psi}_
{y1,y2}$ as a function of the interaction time $\kappa{t}$. After the interaction channel is closed, the state of the two
vibrational modes is stationary.}
\label{impulsi}
\end{figure}

The time integration of the effective coupling rate
range where the amplitudes of the laser pulses are non-zero
gives $\vartheta\simeq\pi/2$, so we expect that, after the
effective interaction mediated by the cavity mode, the $y1$
motional state has been transferred to $y2$. This can be seen
plotting the fidelity ${\cal
F}_{\psi}(t)=_{y1,y2}\!\bra{0,\psi}{\$(\rho_{\psi})}\ket{0,\psi}_{y1,y2}$,
as a function of $\kappa{t}$. Here, $\$$ is the map given by the
reduced master equation we have derived. It takes the initial
density matrix $\rho_{\psi}=\ket{\psi,0}_{y1,y2}\!\bra{\psi,0}$
into $\rho_{v}(t)$, the solution of Eq.~(\ref{ME1}). The fidelity
turns out to be a function of the interaction time and is parametrized
by the state we want to transfer. The results of our simulation are
presented in Fig.~\ref{impulsi} {\bf (b)}. The fidelity is very
good, reaching $\simeq0.9$ for $t\gtrsim200\kappa^{-1}$.
We have plotted ${\cal F}_{\psi}(t)$ for interaction times larger than
$\mathbf T$ to show that, once the effective coupling is turned off, the
interaction channel breaks down and the state of the ions
becomes stationary. It is worth stressing that the fidelity at
the beginning of the interaction is non-zero because of the presence of
$\ket{00}_{y1,y2}$ in both the initial and target states.

The second important point that has to be addressed in order to
completely characterize the performance of the state transfer
protocol is the leakage. We can single out two different kinds of
leakage. One kind is to states such as $\ket{i,j}_{y1,y2}$
($i,j\in\{1,3\}$) which are the states of the computational basis
having more than $2$ phonons in the $y2$ mode and some
phonons in the $y1$ mode. The other kind of leakage leads to
states lying outside the computational space. The influence of
both these sources of error can be contemplated looking at the
norm of the final density matrix $\rho_{v}(t)=\$(\rho_{\psi})$.
We have checked that $Tr(\rho_{v}(t))=1$ for all the relevant
interaction times, showing that the influence of highly excited
phononic states in the vibrational modes can be neglected. This
indirectly demonstrates that leakage of the latter kind is
irrelevant and the dynamics of the system is confined in the
computational space we have chosen. On the other hand, by
considering the {\it quasi-norm}
${\sum^{2}_{i=0}}\,_{y1,y2}\!\bra{0i}{\$(\rho_{\psi})}
\ket{0i}_{y1,y2}$, we can check how large the influence of
populated $y1$ states is in the density matrix $\rho_{v}(t)$. This
is shown in Fig.~\ref{normTrans} {\bf (a)}
\begin{figure} [ht]
{\bf (a)}\hskip5.0cm{\bf (b)}
\centerline{\psfig{figure=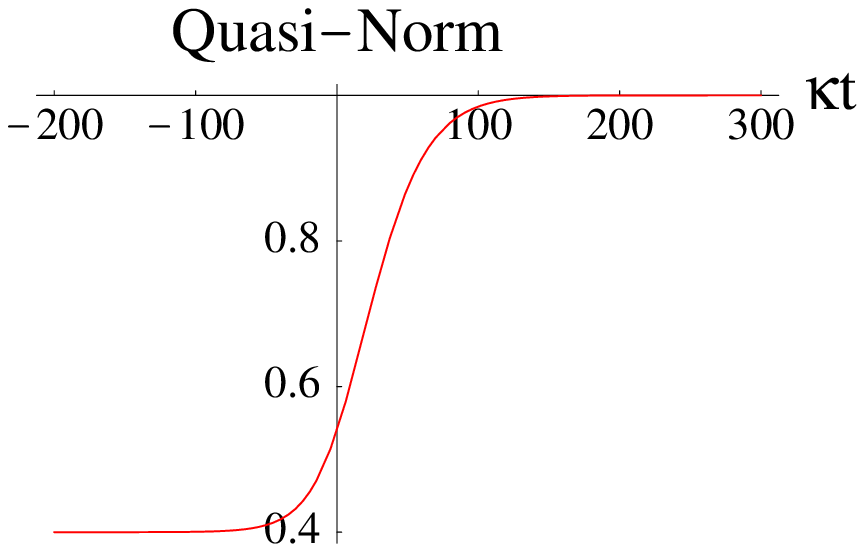,width=4.5cm,height=3.0cm}\psfig{figure=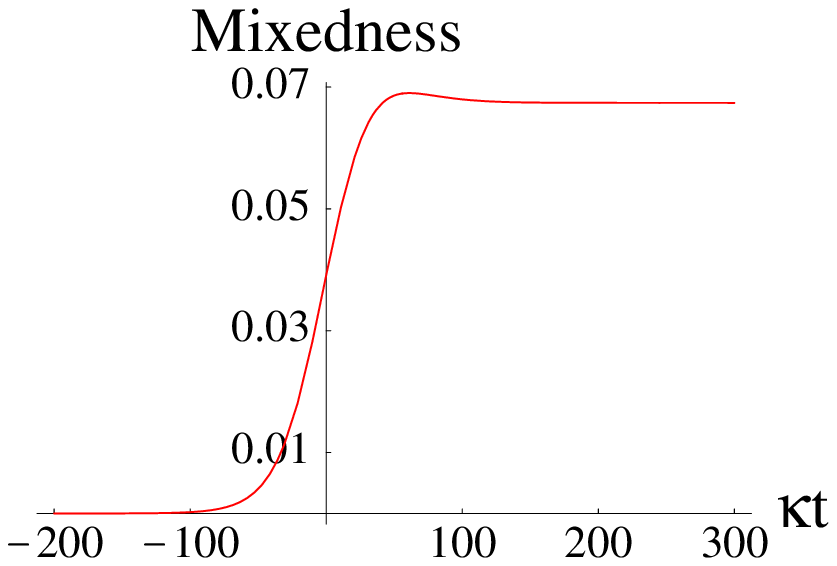,width=4.5cm,height=3.0cm}}
\caption {{\bf (a)}: The quasi-norm
${\sum^{2}_{i=0}}_{y1,y2}\!\sand{0i}{\$(\rho_{\psi})}{0i}_{y1,y2}$
is plotted versus the interaction time $\kappa{t}$. This quantity
considers the contribution to  the normalization of the density
matrix given by the states having just the $y2$ mode populated.
{\bf (b)}: The degree of mixedness of $\rho_{v}(t)$ versus the interaction
time $\kappa{t}$. The purity of the state is measured by the
linearized entropy $S_{L}=(16/15)\left[1-Tr(\rho^{2}_{v})\right]$
that is zero for a perfectly pure state (as the initial one, at
$\kappa{t}=-200$) and $1$ for a statistical mixture.}
\label{normTrans}
\end{figure}
where we can see that, after the transient period when the
contribution by non-empty states of the $y1$ mode is relevant,
the steady state of the system is perfectly normalized. This
means that the final state of the two vibrational modes does not
contain excitations of the {\it transmitter}. To complete this analysis,
we give some insight about the purity of the state we get. From
this viewpoint, the fidelity is not a good tool because it could
give the same quantitative results for $\rho_{v}(t)$ describing a
pure state or a statistical mixture. On the other hand, an easily
computable quantity is the  {\it linearized
entropy}  $S_{L}(t)=(16/15)\left[1-Tr(\rho^{2}_{v}(t))\right]$~
\cite{commento2}. This quantity is zero for a pure state and
reaches $1$ if the state is completely mixed. We show a plot of $S_{L}(t)$
in Fig.~\ref{normTrans} {\bf (b)}. The state remains highly
pure all along the interaction, the small degree of mixedness being due to
the dissipative nature of the effective evolution of the system
arising from the bad-cavity regime. 

The fidelity and purity of the state is not perfect because of the losses
induced on the two-mode vibrational subsystem by the dissipative
bus. However, the state we get with this protocol is nearly
optimal. Higher quality factors of the optical cavity coupled to
the ion traps will improve the performances. In this case,
indeed, we would be able to neglect the dissipative dynamics of
the cavity mode, making the effective interaction between
transmitter and receiver perfectly unitary. The form of the
effective Hamiltonian, arising from the coupling scheme, will be
as in Eq.~(\ref{HamCavRWA}) with the replacement
$\kappa\rightarrow\delta_{c}$ as, in this case, the
condition $\delta_{c}\gg\kappa$ has to be used. The fidelity of
the operation and the purity of the final state will be ideal. 
Unfortunately, this is not the
case of the experiments performed in~\cite{schmidt-kaler}, where
the cavity decay-rate is $\kappa\simeq{2\pi}\times102\,KHz$ and
has to be contrasted to the ion-cavity coupling
$g_{0}\simeq2\pi\times{134}Hz$. These cavity parameters, an
external laser-ion Rabi frequency ${\cal E}_{L}\simeq{30}\,KHz$
with Lamb-Dicke parameter $\eta_{y}\simeq0.2$ and detuning
$\Delta\simeq{50}\,KHz$ (larger than a typical laser bandwidth
$\sim6{KHz}$) allow us to get an effective coupling rate
$\tilde\Gamma\simeq0.03$. This, in turn, gives interaction times
in the range of $0.1\,msec$, smaller than the effective lifetime
of the ion's excited state. It is noticeable that we do not 
require a high-quality factor cavity to achieve an
almost perfect motional state transfer, a feature of the protocol 
we have proposed that is certainly important from the practical
viewpoint.
\section{Quasi-Bell states measurement}
\label{QBD}
We next consider the Bell-state measurement needed
in the protocol for coherent quantum
computation~\cite{jacobmyung}. As we will see in
Section~\ref{twoqubits}, Bell-measurements can be used to
construct the teleportation-based ${\sf CNOT}$ suggested by
Gottesman and Chuang~\cite{gottesman}. In our specific case, the
quantum channel for the teleportation protocol is embodied by one
of the ECS's in Eq.~(\ref{canoniciecs}). For sufficiently large 
amplitudes of their components, the ECS are quasi-orthogonal, carry exactly
one ebit of entanglement and are usually referred to as
quasi-Bell states. A complete discrimination of the elements of
this class is, thus, fundamental in our scheme. It is worth
stressing here the well-known  {\it no-go theorem} demonstrating
that a never-failing, full Bell-state analyzer can not be
realized using just linear interactions ({\it e.g.} beam splitters and
phase shifters)~\cite{norbert}. More recently, it has been
recognized that the introduction of a Kerr non-linear
interaction~\cite{vitalitombesi} or the exploitation of
additional degrees of freedom of the system employed~\cite{beennaker} 
can be used to fully discriminate all four Bell states. However, these schemes
are designed to work with two-level systems and are not relevant
to the infinite dimensional case we treat.

The direct detection of the properties of a vibrational
state is, in general, a hard task to accomplish. On the other
hand, detecting the electronic state of an ion (or an array of
ions) is more straightforward and can be performed using the quantum jump
technique, in which resonance fluorescence from a strongly driven atomic
transition is detected~\cite{wineland,sasura,lieb,blattzoller}.
The presence/absence of fluorescence in
the driven transition reveals the electronic state of the ion.
Thus, we need a joint interaction that changes the internal
degrees of freedom of the ion in a way that reflects the state of
the vibrational ones. The measurement of the electronic state of the
ion after the interaction, then, will give information on its
vibrational state. To achieve this goal, we start by considering 
the Hamiltonian obtained by applying a standing-wave laser
field to the trapped ion. In a rotating frame and with
$\Delta=\omega_{eg}-\omega_{L}$ the detuning between the
standing-wave and the ion's transition frequency, the
interaction reads
\begin{equation}
\label{hamqnd2}
H'={\Delta}\hat{\sigma}_{z}+{\omega_{x}}
\hat{b}^{\dag}_{x}\hat{b}_{x}+\Omega\cos[\eta(\hat{b}^{\dag}_{x}+\hat{b}_{x})]
(\hat{\sigma}_{-}+\hat{\sigma}_{+}),
\end{equation}
where $\Omega$ is the corresponding Rabi frequency. In the
dispersive limit $\Delta\gg\Omega$, with $\Delta$ well-away from
the resonant vibrational frequency~(see Schneider and Milburn
in~\cite{murao}), we can adiabatically eliminate the excited state
of the ion and expand
$\cos[\eta(\hat{b}^{\dag}_{x}+\hat{b}_{x})]$ in power series,
retaining terms up to the second order in $\eta$ (Lamb-Dicke
limit). In the interaction picture and neglecting terms
oscillating at frequency $\pm{2}\omega_{x}$, we get
\begin{equation}
\label{hamqndend}
H_{qnd}\simeq{2}\frac{\Omega^{2}}{\Delta}\eta^{2}\hat{b}^{\dag}_{x}\hat{b}_{x}\hat{\sigma}_{z},
\end{equation}
where state-independent energy terms have been omitted. This Hamiltonian
is suitable for quantum non-demolition
measurements of the motional even/odd coherent states
discussed above. Indeed, the evolution operator
$\hat{U}_{qnd}(t)=\exp\left(-iH_{qnd}t\right)$ does not change 
the parity of an even/odd coherent state but phase-shifts the electronic
state by an amount depending on the vibrational state
phonon-number. Explicitly:
\begin{equation}
\label{Uqnd}
\begin{split}
\hat{U}_{qnd}(t)&=
\cos(\chi_{qnd}t\hat{b}^{\dag}_{x}\hat{b}_{x})\one-i\sin(\chi_{qnd}t\hat{b}^{\dag}_{x}\hat{b}_{x})\hat{\sigma}_{z}\\
&=e^{i\chi_{qnd}t\hat{b}^{\dag}_{x}\hat{b}_{x}}\pro{g}{g}+e^{-i\chi_{qnd}t\hat{b}^{\dag}_{x}\hat{b}_{x}}\pro{e}{e},
\end{split}
\end{equation}
with $\chi_{qnd}=2\Omega^2\eta^2/\Delta$. If we set
$\chi_{qnd}t=\pi/2$, the electronic states will be mutually
shifted $\pi$-out-of-phase. 

Now, let us assume that we prepare
the $x$ vibrational mode of an ion in an even/odd coherent state
${\cal N}_{\pm}(\ket{\alpha}\pm\ket{-\alpha})$, while its
internal state is $\ket{g}$. Then, we apply a $\pi/2$-pulse tuned
on the carrier frequency of the ion's spectrum. This particular
interaction realizes the Hamiltonian $H_{car}={\cal
G}\hat{\sigma}_{+}+h.c.$ that couples
$\ket{gn}\leftrightarrow\ket{en}$ (${\cal
G}$ being a Rabi frequency). That is, it does not affect
the vibrational state~\cite{wineland}. The $\pi/2$-pulse prepares
the superposition $(1/\sqrt{2})\left(\ket{e}+\ket{g}\right)$. The
standing-wave described above is then applied and the interaction
lasts for $t=\pi/2\chi_{qnd}$. This step of the protocol is used
to  {\it write} the vibrational state on the internal degrees of
freedom of the ion. Another carrier-frequency $\pi/2$-pulse mixes
up the phase-shifted components of the electronic state and,
finally, the internal state detection is performed via quantum
jumps. It is worth stressing that the electronic state detection
is a true projective measurement (in the Von Neumann sense) that
is able to tell us if the ion was in $\ket{g}$ or not. In this
latter case, the vibrational state is reconstructed depending on
the outcome of this last step. The described protocol realizes
the transformations
\begin{equation}
\label{evenodd}
\begin{split}
{\cal N}_{+}\left(\ket{\alpha}+\ket{-\alpha}\right)\ket{g}\rightarrow\left(\ket{i\alpha}+\ket{-i\alpha}\right)\ket{g},\\
{\cal N}_{-}\left(\ket{\alpha}-\ket{-\alpha}\right)\ket{g}\rightarrow\left(\ket{i\alpha}-\ket{-i\alpha}\right)\ket{e}.
\end{split}
\end{equation}
Thus, the different parity of the two vibrational states affects
differently the interference between the components of the
Fourier-transformed state $\ket{g}$. The discrimination between
even and odd coherent states can be performed with, in principle,
high accuracy. Each step in the protocol can be, indeed,
quite precisely performed if a judicious choice of the parameters
is made. The preparation of the electronic state superposition
can be done {\it off-line}, exploiting one of the two laser beams
that build up the standing-wave and reminding that the effective
Hamiltonian in Eq.~(\ref{eq2}) does not affect the electronic variables of the
ion (the manipulation of the electronic state then has no influence 
on the vibrational states).
An estimate of the interaction time required to perform
$\hat{U}(\pi/2\chi_{qnd})$ and to achieve the right phase shift leads to
$\sim15\,\mu{sec}$ for $\Omega=\pi\times{500}\,KHz$,
$\eta=0.2$ and $\Delta=10\,MHz$~\cite{myungscheme}.

This protocol, which was studied for the cavity quantum electrodynamic
model to detect even and odd parities of the cavity
field~\cite{englert}, is useful for the detection scheme for ECS.  In
particular, let us suppose an ECS state of the $x$ and $y$ modes of
an ion is subject to a $50:50$ BS operation. 
This will give us one of the output
modes in an even/odd coherent state, the other being in its
vacuum. In particular
\begin{equation}
\begin{split}
\hat{B}_{xy}\left(\frac{\pi}{4}\right) \ket{\phi_{\pm}}_{xy}&=N^{\phi}_{\pm}
\left\{\ket{\sqrt{2}\alpha}\pm\ket{-\sqrt{2}\alpha}\right\}_{x}\otimes\ket{0}_{y},\\
\hat{B}_{xy}\left(\frac{\pi}{4}\right)\ket{\psi_{\pm}}_{xy}&=N^{\psi}_{\pm}
\ket{0}_{x}\otimes\left\{\ket{\sqrt{2}\alpha}\pm\ket{-\sqrt{2}\alpha}\right\}_{y},\\
\end{split}
\end{equation}
where $\hat{B}_{xy}(\pi/4)$ is the $50:50$ BS
operator~\cite{burnett}. Then, the following protocol could be
used. We prepare the electronic state of the trapped ion in the
ground state and apply a carrier-frequency $\pi/2$-pulse to get
the electronic superposition
$\left(\ket{e}+\ket{g}\right)/\sqrt{2}$. Then, we arrange the
evolution $\hat{U}_{qnd}(t/2\chi_{qnd})$ for the 
electronic+$x$-vibrational subsystem. Another carrier-frequency
$\pi/2$-pulse on the ion mixes the components of the electronic
superpositions. A quantum-jump detection reveals the internal
state and the output is recorded. If the result of the
measurement is $\ket{g}$, the entire protocol is re-applied, this
time arranging the $\hat{U}_{qnd}(t/2\chi_{qnd})$ evolution of the 
electronic+$y$-vibrational subsystem. If the result of the first
electronic detection is instead $\ket{e}$, we use a $\pi$-carrier
pulse that restores $\ket{g}$, before the protocol is re-applied
(this can be done with, in principle, $100\%$ of
accuracy~\cite{wineland}). The different combinations in which the
internal state of the ion is found allows us for a partial
discrimination between the elements in the ECS class. 
Denoting $|even,i\sqrt{2}\alpha\rangle_{j}$ and
$|odd,i\sqrt{2}\alpha\rangle_{j}$ ($j=x,y$), respectively, the
even and odd coherent states whose components have absolute
amplitude $\modul{i\sqrt{2}\alpha}$, one can prove the
correspondences shown in Tab.~\ref{tabella1}.
\begin{table} [ht]
\begin{ruledtabular}
\begin{tabular}{|l|r|c|r|}\hline
Initial state&$1^{st}$ det.&$2^{nd}$ det.& Final vibrational state \\ \hline\hline
$\ket{\phi_{+},g}_{xy,ion}$ & $\ket{g}_{ion}$    & $\ket{g}_{ion}$       & $\ket{even,i\sqrt{2}\alpha}_{x}\otimes\ket{0}_{y}$  \\ \hline
$\ket{\phi_{-},g}_{xy,ion}$ & $\ket{e}_{ion}$    & $\ket{g}_{ion}$       & $\ket{odd,i\sqrt{2}\alpha}_{x}\otimes\ket{0}_{y}$  \\ \hline
$\ket{\psi_{+},g}_{xy,ion}$ & $\ket{g}_{ion}$    & $\ket{g}_{ion}$       & $\ket{0}_{x}\otimes\ket{even,i\sqrt{2}\alpha}_{y}$  \\ \hline
$\ket{\psi_{-},g}_{xy,ion}$ & $\ket{g}_{ion}$    & $\ket{e}_{ion}$       & $\ket{0}_{x}\otimes\ket{odd,i\sqrt{2}\alpha}_{y}$  \\ \hline
\end{tabular}
\end{ruledtabular}
\caption{We schematically present  the protocol
for a partial quasi-Bell states discrimination. The initial state
(the vibrational $x$ and $y$ modes being prepared in an ECS), the
outcomes of the $1^{st}$ and $2^{nd}$ electronic measurement and
the final vibrational state are displayed. \label{tabella1}}
\end{table}

It is seen that, while the discrimination between $\ket{\phi_{-}}_{xy},\,\ket{\psi_{-}}_{xy}$ and
$\left\{\ket{\phi_{+}},\ket{\psi_{+}}\right\}_{xy}$ is
perfect, this is not the case for the elements of
the subset $\left\{\ket{\phi_{+}},\ket{\psi_{+}}\right\}_{xy}$.
The sequence of detected electronic measurements corresponding to these vibrational
states is the same and there is no way to distinguish between
them, following this strategy. However, one can exploit the parity non-demolition nature of the
above procedure. Even if the amplitudes of the components
of an even/odd coherent state are changed (the amplitude transforming 
from $\modul{\alpha}$ to $\modul{i\sqrt{2}\alpha}$), the parity eigenvalue of
these states is preserved.

Now, if $\ket{\phi_{+}}_{xy}$ is prepared instead of
$\ket{\psi_{+}}_{xy}$, we end up with mode $x$ being populated
while $y$ is in its vibrational vacuum state. The configuration will be 
contrary if $\ket{\psi_{+}}_{xy}$ is prepared. 
Thus, the key of our procedure is the
discrimination between $\ket{0}_{x}$ and $\ket{even,i\sqrt{2}\alpha}_{}x$. 
Let us suppose that, after the application of the previous
protocol and having found a sequence of two ground states as a result
of the detection procedure (with the prepared vibrational state being totally unknown), we apply the
displacement operator $\hat{D}(-\epsilon)$ to mode $x$, $\epsilon$ being a proper amplitude. If $\ket{\phi_{+}}_{xy}$ was the
initial state, the displacement transforms the state of the $x$
mode into $\left(e^{i2\sqrt{2}\alpha\epsilon}\ket{i\sqrt{2}\alpha-\epsilon}
+e^{-i2\sqrt{2}\alpha\epsilon}\ket{-i\sqrt{2}\alpha-\epsilon}\right)_{x}$.
Taking $2\sqrt{2}\alpha\epsilon=\pi/2$ and $\alpha=2$, then
we get $\epsilon\simeq{0.27}\ll\alpha$. With this angle of
rotation {\it the even coherent state is changed into an approximation of an odd one}. We have flipped the parity of the state. Applying now the quasi-Bell
state detection protocol, as described above, the outcome of the first atomic
measurement becomes $\ket{e}_{ion}$. Does it help in distinguishing between
$\ket{\phi_{+}}_{xy}$ and $\ket{\psi_{+}}_{x1x2}$? The effect of displacement on
the state
$\ket{0}_{x}\otimes\left(\ket{i\sqrt{2}\alpha}+\ket{-i\sqrt{2}\alpha}\right)_{y}$
(that is the final vibrational state if $\ket{\psi_{+}}_{xy}$ is
prepared, as shown in Tab.~\ref{tabella1}) is
$\ket{0}_{x}\rightarrow\ket{-\epsilon}_{x}$. For the amplitudes
of the coherent state $\alpha$ and the angle of rotation
$\epsilon$ chosen above, however, it is
$\ket{-\epsilon}_{x}\simeq0.96\ket{0}_{x}-0.27\ket{1}_{x}$.
Applying again the ECS detection scheme, just before the first electronic detection, we get
\begin{equation}
\label{cdiscr}
\begin{split}
&\ket{-\epsilon}_{x}\otimes\ket{even,i\sqrt{2}\alpha}_{y}\otimes
\ket{g}_{ion}\\&\rightarrow(0.96\ket{0}\ket{g}-i0.27\ket{1}\ket{e})_{x,ion}\otimes\ket{even,i\sqrt{2}\alpha}_{y}.
\end{split}
\end{equation}
The probability to get the electronic output $\ket{e}_{ion}$  is
given by $(0.27)^{2}\ll(0.96)^{2}$. That is, most of the times
(approximately $92\%$ of the times) we will obtain
$\ket{g}_{ion}$, making the discrimination between the two states
complete. This can be taken as an estimate of the efficiency of
the quasi-Bell state measurement because, in the present case,
the efficiency of the {\it detector} apparatus (that is the
efficiency of the quantum jumps technique for electronic state's
detection) can be taken, nominally, $100\%$~\cite{blattzoller}. Our
protocol for quasi-Bell state measurements can be adapted to the
case of two distinct vibrational modes relative to remote trapped
ions. It could be re-designed, {\it mutatis mutandis}, using two
ions and their $x$ motional modes. In this case, the internal
degrees of freedom have to be detected {\it in parallel} and not
{\it sequentially} and the discrimination will be based on the
comparison between different combinations of them. Considering
all the operations to be performed before the ion's
internal-state detection, the overall time required for a
complete discrimination of ECS should be in the range of hundreds
of $\mu{sec}$, which is within the coherence times of the system.

\section{Controlled two-qubit gates}
\label{twoqubits} 
In this Section, we consider controlled two-qubit gates to complete 
our discussion on a possibility of qubit operations using coherent
states of vibrational modes of ions.
As we have remarked in the previous Section, a
quasi-Bell state detection (both {\it local} and {\it
distributed}) is possible. The teleportation-based scheme for a {\sf
CNOT}, indeed, exploits the Bell state measurements to perform
two different steps.  We follow the scheme proposed
in~\cite{jacobmyung} and consider two three-mode GHZ states,
$\ket{\xi}_{a0,a1,a2}$ and $\ket{\xi}_{a3,a4,a5}$, of general bosonic modes
$a0,..,a5$. The joint state of modes $a0$ and $a3$ is first
projected onto the Bell basis in order to prepare the (un-normalized) four-mode
entangled state
\begin{equation}
\label{fourmodeancillary}
\begin{split}
\ket{\eta}_{anc}=\ket{\alpha,\alpha}_{a1,a2}\ket{\phi_{+}}_{a4,a5}+\ket{-\alpha,-\alpha}_{a1,a2}\ket{\psi_{+}}_{a4,a5}.
\end{split}
\end{equation}
This is then used to realize the ${\sf CNOT}$ gate as described in refs.~\cite{jacobmyung,gottesman}. The GHZ states can be
built by using beam splitters~\cite{STK} and
single-mode rotations as those already demonstrated above. It has been
recently recognized (see for example~\cite{ralph}) that the
four-mode state, being a complicated step to perform, can be prepared {\it off-line} and
then used in the protocol for the {\sf CNOT} only when it is needed.
Overall, we need $8$ vibrational modes to
implement a single {\sf CNOT}, six of which are used to prepare for
$\ket{\eta}_{anc}$ and the remaining two for the control and
target qubits. In our scheme, however, we use two vibrational modes per ion so that
we require $4$ ions and many motional
state transfer operations. Even if there is no in-principle
difficulty in doing this, it is apparent that the scheme is
experimentally challenging, not just for the {\it in situ}
operations to perform (linear and non-linear coupling between
orthogonal vibrational modes) but for the transfer protocol (that is 
the slow and less efficient part of our
scheme). However, it is straightforward to extend the Hamiltonian
model in Eq.~(\ref{eq2}) to three orthogonal vibrational modes
(that is to include the $z$ mode in the coupling model)
using laser fields having (opposite) projection onto the azimuthal axis
too. In this way, we will be able to exploit three modes per ion,
altogether, and the steps necessary to create $\ket{\eta}_{anc}$
can be performed using a two-ion crystal in the optical cavity.
The projected modes $a0$ and $a3$, having not been involved in the
four-mode ancillary state, could be used to embody the target and
control of a two-qubit gate. A single {\sf CNOT} gate, thus, can be
realized with just a two-element register.

Furthermore, our ability to engineer the Kerr non-linearity can be used here
to reduce the number of teleporting operations we have to perform. We
exploit that 
$\ket{\phi_{+}}_{ai,aj}$ and $\ket{\psi_{+}}_{ai,aj}$ are
mutually swapped by the {\it cross-parity} operator
$(-1)^{\hat{n}_{ai}\hat{n}_{aj}}$. On the other hand,
$\ket{\phi_{-}}_{ai,aj}$ and $\ket{\psi_{-}}_{ai,aj}$ are not
affected by this evolution. Schematically:
\begin{equation}
\label{ECSprop}
\begin{split}
(-1)^{\hat{n}_a\hat{n}_b}\ket{\phi_{+}}_{ab}=\ket{\psi_{+}}_{ab},\\
(-1)^{\hat{n}_a\hat{n}_b}\ket{\psi_{+}}_{ab}=\ket{\phi_{+}}_{ab},\\
(-1)^{\hat{n}_a\hat{n}_b}\ket{\phi_{-}}_{ab}=\ket{\phi_{-}}_{ab},\\
(-1)^{\hat{n}_a\hat{n}_b}\ket{\psi_{-}}_{ab}=\ket{\psi_{-}}_{ab}.
\end{split}
\end{equation}
The operator
$(-1)^{\hat{n}_a\hat{n}_b}\equiv{e}^{-i\pi{\hat{n}_a\hat{n}_b}}$
is implemented by the cross-phase modulation  used to create the ECS
from separable coherent states, as described in Section \ref{QSE}.
Here, we are interested in the effect of this
unitary operation on the class of ECSs. Now, a cross-phase 
modulation between vibrational modes $a1$
and $a3$ is assumed to be used so that the state
$\ket{\sqrt{2}\alpha}_{a1}\left(\ket{\sqrt{2}\alpha}+
\ket{-\sqrt{2}\alpha}\right)_{a3}+\ket{-\sqrt{2}\alpha}_{a1}
\left(\ket{\sqrt{2}\alpha}-\ket{-\sqrt{2}\alpha}\right)_{a3}$ is
generated~\cite{ioEIT}. We consider the vibrational modes $a2$
and $a4$, each prepared in the vacuum state, and realize a
$50:50$ beam splitting in the $a1+a2$ and $a3+a4$ subsystems.
Finally, the $(-1)^{\hat{n}_{a3}\hat{n}_{a4}}$ produces a state (not normalized) that, with local unitary operations (a single-qubit ${\sf
NOT}$ gate), can be written as
\begin{equation}
\label{channel}
\begin{split}
\ket{\eta{'}}_{anc}=\ket{\alpha,\alpha}_{a1,a2}\ket{\phi_{+}}_{a3,a4}+\ket{-\alpha,-\alpha}_{a1,a2}\ket{\psi_{-}}_{a3,a4}.
\end{split}
\end{equation}
This differs from $\ket{\eta}_{anc}$ because it involves
$\ket{\psi_{-}}_{a3,a4}$ (correlated to
$\ket{-\alpha,-\alpha}_{a1,a2}$) instead of
$\ket{\psi_{+}}_{a3,a4}$. We do not try to reproduce the
four-mode entangled channel $\ket{\eta}_{anc}$~\cite{gottesman} but we
go on with $\ket{\eta{'}}_{anc}$ and apply the protocol for a
teleportation-based two-qubit gate. It can be proved by
inspection that, in this case, a {\it controlled}-$i\hat{\sigma}_{y}$
(${\sf C}_{i\sigma_{y}}$) gate between the control $c$ and the target $t$
qubits is realized (up to single-qubit rotations to be applied,
conditionally on the outcomes of the Bell detections).
In the computational basis
$\left\{\ket{\alpha,\alpha},\,\ket{\alpha,-\alpha},\,\ket{-\alpha,\alpha},\,\ket{-\alpha,-\alpha}\right\}_{ct}$,
this can is represented by the block diagonal matrix
${\sf C}_{i\sigma_{y}}=\text{diag}[\one,i\bf{\sigma_{y}}]$,
with ${\bf \one}$ the $2\times{2}$ identity matrix and 
${\bf \sigma_{y}}$ the $y$-Pauli matrix. This gate is 
non-local and is not locally-equivalent to a
${\sf SWAP}$ gate. It is indeed easy to see that
${\sf C}_{i\sigma_{y}}$ is an entangling gate that transforms the
separable state
$(A\ket{\alpha}+B\ket{-\alpha})_{c}\otimes(C\ket{\alpha}+D\ket{-\alpha})_{t}$
into the entangled state
$A\ket{\alpha}_{c}(C\ket{\alpha}+D\ket{-\alpha})_{t}+B\ket{-\alpha}_{c}(D\ket{\alpha}-C\ket{-\alpha})_{t}$.
Thus, together with the single-qubit rotations in Section II, 
this two qubit operation can be used to perform the universal quantum
computation~\cite{bremner}. Moreover, using the
criteria of refs.~\cite{whaley}, this gate turns out to be
locally equivalent to ${\sf CNOT}$.

On the other hand, an important simplification in the realization
of  a two-qubit gate can be achieved if we renounce the four-mode
entangled channel $\ket{\eta}_{anc}$ (or $\ket{\eta{'}}_{anc}$).
This latter can be replaced by two $\ket{\phi_{+}}$ ECS states
that are used as quantum channels for the teleportations of the
output modes of a BS (reflectivity
$\cos^{2}(\theta/2)$) that has superimposed the control and target
qubits (see Fig.~\ref{cnot}). For proper choices of
$\theta=\pi/4\alpha^2$, after the beam splitter operation and the
two teleportations, the output modes ($x1$ and $x2$ in
Fig.~\ref{cnot}) are in a state that is equivalent to
$\cnot$, up to single-qubit operations.
To achieve this result with a significant probability of success,
however, the condition $\theta^{2}\alpha^2\ll{1}$ has to be
fulfilled. For example, if we take $\alpha=2$, 
$\theta=\pi/36$ has to be taken (corresponding to an
interaction time with the external laser fields of some~${\mu{sec}}$, 
for the parameters used in this paper) and
giving a probability of success $\simeq0.92$.
\begin{figure}
\centerline{\psfig{figure=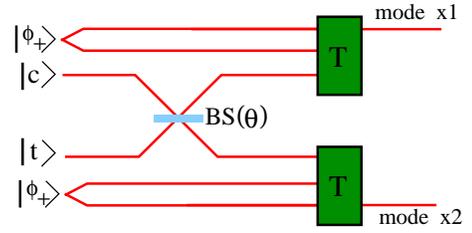,width=6.0cm,height=3.0cm}}
\caption {Scheme for a coherent-state {\sf CNOT}. Target ($\ket{t}$) and control qubit ($\ket{c}$) are superimposed at a
beam splitter (reflectivity $\cos^{2}(\theta/2)$). The states of the 
output modes are, then, teleported onto the state of
mode $x1$ and $x2$. The boxes labeled $T$ resume the teleportation operations.}
\label{cnot}
\end{figure}
Explicitly, the control and target qubits can be {\it written on}
the $y$ modes of the trapped ions $1$ and $2$. The ECSs we need
for the teleportations are codified in the state of the
$x$ and $z$ modes. The beam splitter operation between the control
and target qubits is, in this case, the sole {\it delocalized} operation
we need to perform. The scheme we have described for
motional-state transfer is exactly what we need to split the
input modes in a distributed way. All the other
operations do not involve coupling between the ions of the
register. The Bell-state measurement is finally performed involving $z$
and $y$ motional modes, projecting the $x$ modes onto a state
equivalent to $\cnot\ket{c,t}_{x1,x2}$.
We conclude our analysis with a remark concerning the strategy to follow in order to discriminate the logical states of the qubit (namely between $\ket{\alpha}$ and $\ket{-\alpha}$). This can be done {\it locally}, following ref.~\cite{jacobmyung}, using a $50:50$ BS operation superimposing the state of the qubit, codified in the $x$ mode of an ion, to a coherent state of the ancillary $y$ mode. Then, with proper resonant transitions coupling the internal state of the ion to the $x$ and $y$ modes and highly-efficient electronic state detection, we can ascertain the state of the qubit~\cite{wineland}. The generalization of this procedure to a {\it delocalized} situation can be done exploiting the results shown in Section~\ref{IC}.

\section{Remarks}
\label{remarks}

In this paper we have presented a scheme that, exploiting  the
motional degrees of freedom of individually trapped ions, could
allow for coherent-state quantum
computation~\cite{jacobmyung,ralph}. We have addressed a model
for quantum engineering based on the use  of two
non-resonant laser pulses. By regulating the direction of the lasers
along the trap axes and tuning their frequencies to excite proper
sidebands, we realize various linear and non-linear interactions,
both for the one-qubit and the two-qubit operations. To scale up the
dimension of a quantum register, we have considered a
distributed design of the quantum computer, each node of the
network being a single trapped ion. The interconnections between
remote nodes are established by a
cavity-bus coupled to the transition of the selected ion
~\cite{schmidt-kaler, walther}. Motional state transfer, in this
way, is shown to be realizable with good fidelity and without the
requirement of a high-quality factor cavity. Finally, an
efficient quasi-Bell state discrimination is possible, in this
set-up, using unitary rotations of the states belonging to the
ECS class and inferring the parity eigenvalues of superposition
of coherent states via high-efficiency electronic detections. The
accuracy of this scheme can be, in principle, arbitrarily near to
$100\%$ due  to the exploitation of the additional degree of
freedom represented by the electronic state of the ion. This
feature allows us to circumvent the bottleneck represented by the
{\it no-go theorem} in ref.~\cite{norbert}. We have addressed the
issues of efficiency and practicality of our proposal showing
that, singularly taken, each step of the scheme is foreseeable
with the current state of the art technology, the main difficulty, up
to date, being represented by the sequential combination of them.


\acknowledgments

This work was supported in part by the European Union, the UK
Engineering and Physical Sciences Research Council and the Korea
Research Foundation (2003-070-C00024). M.P. thanks the International Research Centre for Experimental Physics for financial support.



\begin{thebibliography}{99}
\bibitem{commento} We are considering a quadrupole transition in
which the ion is coupled to the {\it gradient} of the electric field. Thus, the optimal coupling is at the nodes of the field.
\bibitem{commento2} Generally speaking, it is
$S_{L}=\frac{N}{N-1}\left[1-Tr(\rho^{2})\right]$,
where $N$ is the dimension of the Hilbert-Schmidt space
in which $\rho$ is defined.
In our case, $N=4\times{4}$ because of the restriction of
the computational basis.
\bibitem{myungscheme} An alternative way to discriminate
even and odd coherent states requires the
Hadamard gate and a beam splitter operation. Indeed, the application of
a Hadamard gate (for example the one we have discussed in Section~\ref{QSE})
to an even (odd) coherent state produces the coherent state $\ket{\alpha}$
($\ket{-\alpha}$). Mixing this at a $50:50$ beam splitter with another coherent state
of amplitude $\alpha$ ($-\alpha$) produces the beam-splitter's output state
$\ket{\sqrt{2}\alpha}_{out1}\otimes\ket{0}_{out2}$ $\left(\ket{0}_{out1}
\otimes\ket{-\sqrt{2}\alpha}_{out2}\right)$. Thus, the distinction between the
two class of states reduces to the discrimination between a zero-phonon state
and a coherent state of amplitude $\modul{2\alpha}$ in the output mode $out1$,
for example (represented by one of the vibrational modes of the trapped ion) and this can be performed quite efficiently as described in~\cite{wineland}.
\end{thebibliography}
\end{document}